\documentclass[12pt,a4paper,final]{iopart}

\usepackage{iopams}  
\usepackage{graphicx}

\usepackage{amssymb,bm, float, verbatim}
\usepackage[breaklinks=true,colorlinks=true,linkcolor=blue,urlcolor=blue,citecolor=blue]{hyperref}

\begin{document}

\title[Quantum radiation reaction in head-on laser-electron beam interaction]{Quantum radiation reaction in head-on laser-electron beam interaction}

\author[cor1]{Marija Vranic}
\address{GoLP/Instituto de Plasmas e Fus\~ao Nuclear, Instituto Superior T\'ecnico, Universidade de Lisboa, 1049-001 Lisbon, Portugal}
\ead{marija.vranic@ist.utl.pt}

\author{Thomas Grismayer}
\address{GoLP/Instituto de Plasmas e Fus\~ao Nuclear, Instituto Superior T\'ecnico, Universidade de Lisboa, 1049-001 Lisbon, Portugal}
\ead{thomas.grismayer@ist.utl.pt}

\author{Ricardo A. Fonseca$^{1,2}$}
\address{$^1$GoLP/Instituto de Plasmas e Fus\~ao Nuclear, Instituto Superior T\'ecnico, Universidade de Lisboa, 1049-001 Lisbon, Portugal}
\address{$^2$DCTI/ISCTE - Instituto Universit\'ario de Lisboa, 1649-026 Lisboa, Portugal}
\ead{ricardo.fonseca@ist.utl.pt}

\author{Luis O. Silva}
\address{GoLP/Instituto de Plasmas e Fus\~ao Nuclear, Instituto Superior T\'ecnico, Universidade de Lisboa, 1049-001 Lisbon, Portugal}
\ead{luis.silva@ist.utl.pt}

\begin{abstract}
In this paper, we investigate the evolution of the energy spread and the divergence of electron beams while they interact with different laser pulses at intensities where quantum effects and radiation reaction are of relevance. The interaction is modelled with a QED-PIC code and the results are compared with those obtained using a standard PIC code with a classical radiation reaction module. In addition, an analytical model is presented that estimates the value of the final electron energy spread after the interaction with the laser has finished. While classical radiation reaction is a continuous process, in QED, radiation emission is stochastic. The two pictures reconcile in the limit when the emitted photons energy is small compared to the energy of the emitting electrons. The energy spread of the electron distribution function always tends to decrease with classical radiation reaction, whereas the stochastic QED emission can also enlarge it. These two tendencies compete in the QED-dominated regime. Our analysis, supported by the QED module, reveals an upper limit to the maximal attainable energy spread due to stochasticity that depends on laser intensity and the electron beam average energy. Beyond this limit, the energy spread decreases.  These findings are verified for different laser pulse lengths ranging from short $\sim$ 30 fs pulses presently available to the long $\sim$ 150 fs pulses expected in the near-future laser facilities, and compared with a theoretical model. Our results also show that near future experiments will be able to probe this transition and to demonstrate the competition between enhanced QED induced energy spread and energy spectrum narrowing from classical radiation reaction.

\end{abstract}

\pacs{52.65.Rr, 41.75.Ht, 11.80.-m, 52.65.Cc}
\vspace{2pc}
\noindent{\it Keywords}: particle-in-cell, classical radiation reaction, quantum radiation reaction, laser-electron interaction,
\submitto{\NJP}

\section{Introduction}

Near future facilities \cite{ELI, ELI_design, HiPER} will provide extreme laser intensities ($I>10^{22}~\mathrm{W/cm^2}$), where quantum effects such as electron-positron pair production and discrete photon emission might play a central role in laser-matter interaction \cite{Gremillet, Ridgers_quantumRRnew, Sarri_ep, Bulanov_multiple_lasers, Yoffe_evolution, Elkina_rot, Ridgers_solid, Nerush_laserlimit, Bell_Kirk_MC}. Previously, electron-positron pairs have been produced in experiments using a moderately-intense laser of intensity $I\sim 10^{18}~\mathrm{W/cm^2}$ counter-propagating with the ultra-relativistic (46 GeV) SLAC electron beam \cite{QED1,QED2, QED3}. This setup takes advantage of the ultra-relativistic energy of the particles to observe certain nonlinear quantum effects in electromagnetic fields whose amplitude remains several orders of magnitude below the critical Schwinger field \cite{Schwinger}, which constitutes usually the threshold to observe pairs spontaneously created in vacuum. As detailed in \cite{Ritus_thesis}, the field magnitude in the rest frame of the particles will then be of the order of the critical field and the probability of the process becomes optimal. By leveraging on the tremendous progress accomplished in laser technology in the last decades, one can also envisage nowadays to decrease the energy of the relativistic particles and increase proportionally the magnitude of the field of the laser. This explains the recent growing interest \cite{Yoffe_evolution, kogaPOP, capturePiazza, ThomasRR, Keita_neweq, RRinCascade_classical, Ruhl_threshold} on configurations where a relativistic electron beam interacts with laser pulses of significantly higher intensity than in the SLAC experiment. The typical electron energy sufficient to diagnose nonlinear quantum effects is around a few GeV and such electron beams can now be generated from an all-optical source in an efficient manner: the current experimental record for self-injected electrons obtained in a laser wakefield accelerator is 4 GeV \cite{Leemans_4gev}. This shows that the near future laser facilities can be used to explore this nonlinear Compton scattering configuration without the aid of conventional accelerators. 

In our previous work \cite{First_PRL} we have studied the radiation reaction for an electron beam in a laser field with the use of the Landau-Lifshitz equation \cite{LLbook} which has been recognised as the best candidate to describe classically the effect of radiation reaction on charged particle orbits \cite{LLfromQEDsoviet, spohn,Spohnbook,Piazza_solutionLL,qed_class_rr, Vranic_ClassicalRR}. In the Landau-Lifshitz equation framework, charged particles 
 emit radiation continuously and the direct effect of this emission can be represented as a continuous drag force in the particle motion equation. As shown in \cite{First_PRL}, when a GeV electron beam collides head-on with an intense laser ($I\simeq10^{21}~\mathrm{W/cm^2}$), one of the key effects of classical radiation reaction is to reduce the width of the electron energy distribution function during the interaction \cite{Yoffe_evolution,esarey_thompson}. The tendency of classical radiation reaction to shrink the electron energy distribution function has been studied also in fusion plasmas \cite{RR_in_fusion}, and in ion acceleration with solid targets \cite{Tamburini_NIMR}. If the laser intensity is raised such that the classical description of radiation reaction becomes inapplicable, the quantum effects in radiation reaction induce the opposite behaviour, leading to an increase of the energy spread of the beam spectra \cite{Piazza_qed_energyspread}. With the advent of quantum electrodynamics (QED) modules incorporated in the traditional particle-in-cell (PIC) algorithm, we are now able to simulate from first principles quantum radiation reaction in laser-plasma interaction and therefore to validate some of the recent theoretical predictions. This paper thus deals with differences between the classical and the quantum electrodynamics (QED) description in the transition regime where the probabilities for pair creation are still negligible, but quantum effects in the photon emission can already be significant ( "moderately quantum regime" defined in \cite{Piazza_QRR_FD}). This is of particular relevance since upcoming experiments at several facilities will be able to operate in this regime. We carry out PIC-QED simulations that allow us to evaluate the influence of quantum emission on the electron energy spread and the divergence of the electron beam. Maximum attainable energy spread due to quantum stochasticity as a function of mean electron energy and the laser intensity is computed. This result further allows us to obtain a semi-classical analytical prediction for the electron energy spread after the shutdown of the laser as a function of the initial beam and laser parameters. The prediction is in agreement with fully quantum Monte-Carlo PIC simulations. 
 
This paper is structured as follows. In Section 2, we introduce the QED framework to describe photon emission. We then analyse, in Section 3, the evolution of the electron energy spectra, predicted analytically, and we compare the analytical results with QED-PIC simulations. In Section 4, we study the evolution of the electron beam divergence, another measurable quantity in these scenarios, and in Section 5 we state the conclusions of this work. 

\section{QED Photon emission}

In QED, radiation is a discrete stochastic process and this impacts the particle trajectory in a distinct manner from the continuous emission associated with the classical radiation processes. The probabilities of the various processes in an electromagnetic plane wave are based on Volkov states \cite{Volkov}  where the quantum-transition probability is evaluated taking into account the interaction between the particle and the background wave. In the event of emission, there is a transfer of energy from the electron to the emitted photon; otherwise, the electron momentum and energy remain unaltered. The classical limit corresponds to the case where a large number of photons, whose energy remains small compared to the electron energy, is radiated: the high frequency of the emission events allows the approximation of the trajectory as a classical trajectory with a continuous drag. The main difference between the classical and the QED approach is that QED accounts for the possibility of emitting high-energy photons even in a setup where the cross-section for Compton scattering is small (i.e. the average energy loss of the particle is negligible). In the quantum regime, the stochastic nature of emission becomes noticeable and one may expect a diffusion in energy around the mean value of the energy loss, as it was first reported in refs.\cite{Ridgers_quantumRRnew, Piazza_qed_energyspread}. 

The total probability of radiation emission by a single particle is relativistically invariant and depends on the normalised vector potential $a_0=eE/(mc \omega_0)$ and the quantum invariant parameter  $\chi$ ($\chi_e$ for electrons and $\chi_\gamma$ for photons) defined as:
\begin{equation}
\chi_e=\frac{\sqrt{(p_\mu F^{\mu\nu})^2}}{E_{s}~mc}, \quad \chi_\gamma=\frac{\sqrt{(\hbar k_\mu F^{\mu\nu})^2}}{E_{s}~mc},
\end{equation}
where $p_\mu$ is the particle 4-momentum, $k_\mu$ is the photon wave 4-vector, $F^{\mu\nu}$ the electromagnetic tensor, $E_s=m^2c^3/{\hbar c}$ the Schwinger critical field \cite{Schwinger}, $m$ is the electron mass, $e$ is the elementary charge, $c$ is the speed of light and $\omega_0$ is the frequency of the electromagnetic wave. 
The differential probability rate of photon emission by nonlinear Compton scattering is then given \cite{pair_rate1,pair_rate2,pair_rate3, Erber, NikishovRitus, Ritus_thesis} by
\begin{equation}\label{compton_rate}
\frac{d^2 P}{dt~d\chi_\gamma}=\frac{\alpha m c^2}{\sqrt{3}\pi \hbar \gamma \chi_e} 
\left[ \left( 1 - \xi +\frac{1}{1-\xi} \right) K_{2/3}(\tilde{\chi}) - \int^\infty _{\tilde{\chi}} dx K_{1/3}(x)  \right]
\end{equation}
where $\tilde{\chi}=2\xi/(3\chi_e(1-\xi))$,  $\xi=\chi_\gamma/\chi_e$ and $\alpha=e^2/(\hbar c)$ is the fine-structure constant. 

In order to simulate the emission of photons (and electron-positron pairs), we have added a QED module \cite{Thomas_POP_2016, thomasQED} to OSIRIS \cite{OSIRIS} which allows real photon emission from an electron or a positron and decay of the photons into pairs (Breit-Wheeler process). The OSIRIS-QED framework accounts for the differential emission probability rate (\ref{compton_rate}) in a similar fashion as other QED-PIC modules \cite{Elkina_rot, Ridgers_solid, Nerush_laserlimit, Gonoskov_schemesQEDPIC}. The QED algorithm can be summarized as follows: at particle push-time, the probability of radiating a photon is evaluated, and if the event occurs, the radiated photon quantum parameter is selected to obey the distribution given by Eq. (\ref{compton_rate}); the particle momentum is then updated to account for the momentum of the emitted photon (assumed to be radiated in the direction of the particle motion). For Breit-Wheeler pair production, the procedure is similar but instead of emission, we evaluate the probabilities of photon decay into an electron-positron pair. If the event occurs, we then remove the photon and initialize the new particles. The OSIRIS-QED framework is also equipped with an advanced macroparticle merging algorithm \cite{Vranic_merging}. 
\begin{figure}
\centering
\includegraphics[width=0.8\textwidth]{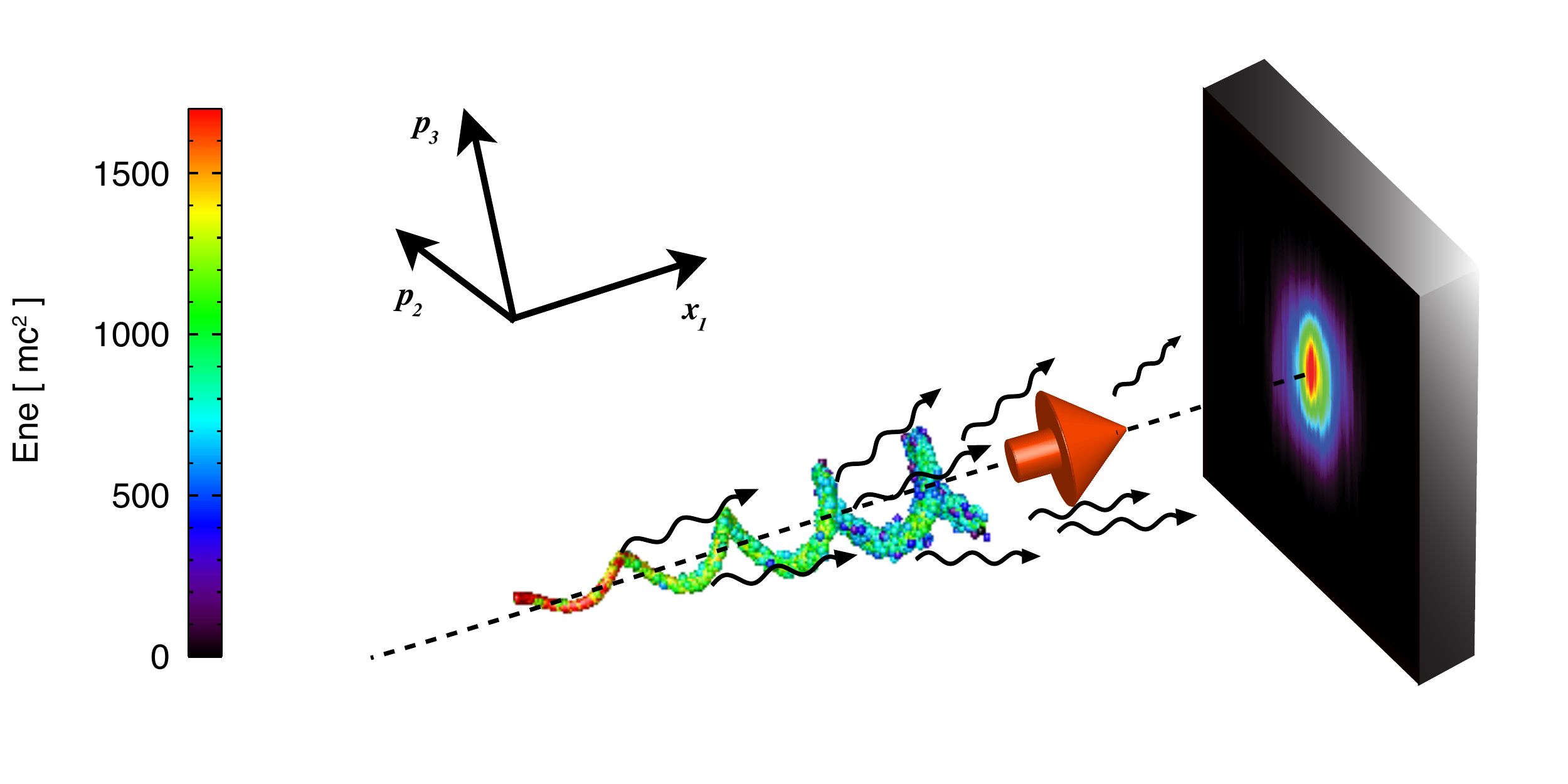}
\caption{QED radiation reaction and photon detection. The electron beam depicted is interacting with a counter-propagating laser  during the rise-time of the temporal laser envelope function.  Individual events of photon emission cause non-continuous energy loss, which is illustrated by the different colours (particle energies) for electrons experiencing the same field.  } 
\label{qedrr_cartoon}
\end{figure}

\section{The evolution of the electron energy spectra}

\begin{figure}[t!]
\centering
\includegraphics[width=0.8\textwidth]{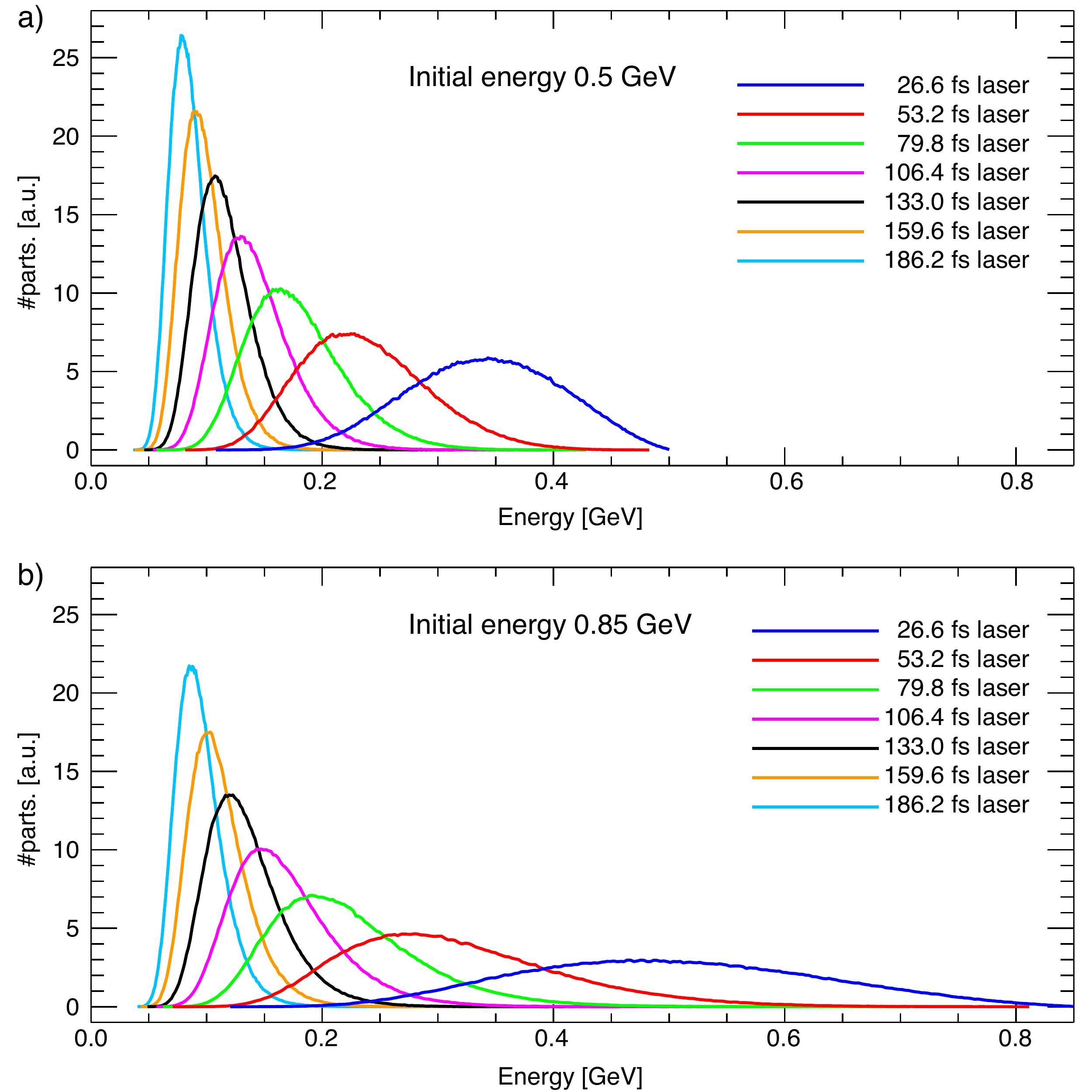}
\caption{Final electron energy spectra: a) starting from the 0.5 GeV electron beam; b) starting from 0.85 GeV electron beam. } 
\label{qedrr_spect}
\end{figure}

We first examine the temporal evolution of the energy spectrum of an electron beam as it collides head-on with an intense laser. In particular, we will focus on how the electron beam energy spread is affected by the QED photon emission. To facilitate this analysis, we define the beam width at a time $t$ as the standard deviation in energy over all the particles 
\begin{equation}
\sigma(t)=\sqrt{\frac{1}{N}\sum_{i=1}^N \left(\overline{\gamma(t)} - \gamma_i(t)\right)^2}
\end{equation}
where $N$ is the total number of particles, $\overline{\gamma(t)}$ is the mean energy of the electron beam at time $t$, as measured in the laboratory frame, and $\gamma_i(t)$ is the energy of the particle $i$ at the same time $t$. The stochastic effects in quantum radiation reaction that are responsible for the spreading of the distribution have been analytically studied in a similar setup in \cite{Piazza_qed_energyspread} where the Fokker-Planck equation is used to describe the evolution of the electron distribution function in time. This approach is valid for $\chi_\gamma \ll 1$. The Fokker-Planck equation \cite{LLphyskin, FP-Planck, FP-Fokker} is usually used in laser plasma interaction, for instance, to model kinetically the collisions between species. One can however see the quantum photon emission as a virtual inelastic collision with an electron; as long as the momentum exchange remains small compared with the emitting particle momentum, the Fokker-Planck equation proves to be adequate.  

If $w(\vec{p},\vec{q})d^3\vec{q}$ denotes the probability per unit time of momentum change $\vec{p}\rightarrow \vec{p}-\vec{q}$ of an electron $\vec{p}$, then the transport equation for the electron distribution function $f(t,\vec{p})$ is given by:
\begin{equation}\label{transport_FP}
\frac{\partial f (t,\vec{p})}{\partial t}=\frac{\partial}{\partial p_l}\left[ A_l f+\frac{1}{2}\frac{\partial}{\partial p_k}(B_{lk}f) \right]
\end{equation}
where 
\begin{equation}
A_l=\int q_l w(\vec{p},\vec{q})d^3\vec{q}, \quad B_{lk}=\int q_l q_k w(\vec{p},\vec{q})d^3\vec{q}
\end{equation}
represent the drift and diffusion coefficients respectively, and indexes $l$ and $k$ denote different spatial components. Under the assumption that the electron beam is relativistic, and that the photons are radiated in the direction of motion, the problem is reduced to one dimension. 

Since the emission probability is given by Eq. (\ref{compton_rate}) as a function of $\chi_\gamma$, we proceed to a change of variables using $\chi_\gamma/\chi_e\approx \hbar k/\gamma mc$ which is a consequence of the collinearity of the electrons and the emitted photons. We then get
\begin{equation}
A=\frac{\gamma mc}{\chi_e} \int_0^{\chi_e}\frac{d^2P}{dt~ d\chi_\gamma}\chi_\gamma d\chi_\gamma, \quad B=\frac{(\gamma mc )^2}{\chi_e^2} \int_0^{\chi_e}\frac{d^2P}{dt~ d\chi_\gamma}\chi_\gamma^2 d\chi_\gamma.
\end{equation}
After integration, the drift coefficient and the diffusion coefficient become respectively 
\begin{equation}\label{Adrift}
A\approx \frac{2}{3} \frac{\alpha m^2 c^3}{ \hbar} \chi_e^2, \quad B\approx\frac{55}{24\sqrt{3}} \frac{\alpha m^3 c^4}{\hbar}~ \gamma~\chi_e^3,
\end{equation}
which were first calculated in \cite{Piazza_qed_energyspread}. The Fokker-Planck equation (\ref{transport_FP}), which is a special case of the master equation in the continuous limit, is valid for $q \ll p$. In our scenario, this validity condition can be expressed as $\chi_{\gamma} \ll \chi_e$ which is only conceivable for $\chi_{e} \ll 1$. 

If there is no diffusion ($B=0$), the equation of the characteristic in Eq. (\ref{transport_FP}) is $dp/dt\simeq mcd\gamma /dt=-A$. For electrons counter-propagating with a linearly polarised wave $\chi_e$ is given by $\chi_e=\sqrt{2} \gamma a_0 \hbar \omega_0/m c^2$, while in a circularly polarised wave $\chi_e=2 \gamma a_0 \hbar \omega_0/m c^2$. This allows us to retrieve the classical result where the photon emission results in the electron relativistic factor $\gamma$ decrease. The rate of this decrease in a linearly polarised wave is given \cite{ ThomasRR, First_PRL} by 
\begin{equation}\label{dgdg_help}
\frac{d\gamma}{dt}=-\alpha_\mathrm{rr} \gamma^2, \quad \alpha_\mathrm{rr}=\frac{4e^2\omega_0^2}{3mc^3}a_0^2
\end{equation}
where $\alpha_\mathrm{rr}$ is a constant with units of frequency. For a circularly polarised wave $\alpha_\mathrm{rr}$ needs to be multiplied by a factor of two. By integrating Eq. (\ref{dgdg_help}) with $\gamma_0$ for initial Lorentz factor, we obtain $\gamma=\gamma_0/(1+\alpha_\mathrm{rr} \gamma_0 t)$ in a linearly polarised wave and $\gamma=\gamma_0/(1+2\alpha_\mathrm{rr} \gamma_0 t)$ in a circularly polarised wave, in agreement with \cite{Piazza_solutionLL}. By neglecting diffusion and assuming an initial Gaussian distribution for the electrons with initial standard deviation $\sigma_0$ and initial mean energy $\overline{\gamma_0}$, the authors in ref. \cite{Piazza_qed_energyspread} have shown that if $\sigma_0 \ll \overline{ \gamma_0}$, the distribution remains approximatively Gaussian with an effective standard deviation
\begin{equation}
\label{sigmaclass}
\sigma(t)=\frac{\sigma_0}{(1+2\alpha_{rr}\overline{\gamma_0} t)^2},
\end{equation}
 which is expressed for a quasi-monoenergetic relativistic electron beam as $\delta \gamma_0/\delta \gamma=(\gamma_0/\gamma)^2$ \cite{First_PRL}. It is not straightforward to rigorously expand this result to account for the diffusion term contribution. However, if now we assume that the drift is negligible (i.e. the average energy remains constant over a period of time $\overline{\gamma}\simeq \overline{\gamma_0}$), we obtain the usual diffusion equation, where we have performed the change of variables $p\simeq mc \gamma$:
\begin{equation}
\frac{\partial f}{\partial t}=\frac{B(t,\overline{ \gamma_0})}{2 m^2c^2} \frac{\partial^2 f}{\partial \gamma^2}.
\end{equation}
In the case of a Gaussian distribution, the standard deviation evolves as
\begin{equation}\label{width_qed_an}
\sigma(t)=\sigma_0\left(1+\frac{1}{\sigma_0^2m^2c^2}\int_0^t B(t')dt'\right)^{1/2}.
\end{equation}
It is therefore clear from Eq. (\ref{sigmaclass}) and Eq. (\ref{width_qed_an}) that there is a competition between the drift-like term that tends to compress the distribution width whereas the diffusion-like term tends to increase it. For an infinitesimally short period of time $dt$, the change of the distribution width at a time $t$ due to the drift is given by differentiating Eq. (\ref{sigmaclass}) yielding
\begin{equation}
d \sigma_1=  - \sigma(t) ~4 \alpha_\mathrm{rr}\overline{\gamma(t)} dt 
\end{equation}
and the change due to the diffusion is obtained in a similar manner by differentiating Eq. (\ref{width_qed_an})
\begin{equation}
d\sigma_2= \sigma(t) \frac{B(t) }{2\sigma(t)^2 m^2c^2}dt.
\end{equation}
We can then compute the total change of the electron distribution width within an interval $dt$:
\begin{equation}\label{realdsig}
d\sigma=\sigma(t)\left[ \frac{B(t) }{2\sigma(t)^2 m^2c^2} - 4\alpha_\mathrm{rr} \overline{\gamma(t)}  \right]dt.
\end{equation}
A direct integration of the Eq. (\ref{realdsig}) is not possible because the variables cannot be separated. The expression from  \cite{Piazza_qed_energyspread} can be retrieved by approximating $\sigma(t)=\sigma_0$ in the term within the squared brackets in Eq. (\ref{realdsig}) and then integrating in time. The authors in \cite{Piazza_qed_energyspread} have shown that their expression is valid for relatively short laser interaction times $\tau$ such that $\alpha a_0 (\overline{ \gamma_0}/\sigma_0)^2\chi_e^2 \tau  \omega_0 \ll 1$. 

Considering that the width of the distribution can change significantly throughout the interaction, it turns out impossible to simplify Eq. (\ref{realdsig}) and obtain an explicit form for the energy spread evolution. Nevertheless, Eq. (\ref{realdsig}) can still provide an insight into the changing features of the electron distribution function. Depending on the specific values of the initial parameters of the beam $d\sigma$ can be positive (the diffusion wins over the drift) or negative (the drift wins over the diffusion). If we start from a narrow momentum spread like in our simulations, the distribution width first tends to increase, and later shrink. The "turning point", when classical-like drift starts winning over the QED-induced diffusion can be defined by solving $d\sigma/dt=0$. For a circularly polarised laser, we obtain 
\begin{equation}\label{eqsigmacp}
\left( \frac{\sigma_T}{\gamma_T}\right)^2\approx\frac{55}{32\sqrt{3}}\frac{\hbar \omega_0}{m c^2}\gamma_T a_0= \frac{2.4}{\lambda_0[\mu \mathrm{m}]}   \times10^{-6}\gamma_T a_0,
\end{equation}
and for a linearly polarised laser,
\begin{equation}\label{eqsigmalp}
\left( \frac{\sigma_T}{\gamma_T}\right)^2\approx\frac{55}{32\sqrt{6}}\frac{\hbar \omega_0}{m c^2}\gamma_T a_0= \frac{1.7}{\lambda_0[\mu \mathrm{m}]}\times10^{-6}\gamma_T a_0 ,
\end{equation}
where $\gamma_T$ and $\sigma_T$ represent the electron average energy and energy spread at the "turning point", $\sigma_T/\gamma_T$ is the relative energy spread and $\lambda_0$ is the laser wavelength. For a given $\gamma_T$ and $a_0$, if $\sigma<\sigma_T$ the energy spread increases, but if $\sigma>\sigma_T$, the width of the energy distribution function decreases. In other words, Eqs. (\ref{eqsigmacp}) and (\ref{eqsigmalp}), allow us to estimate what is the maximum attainable energy spread through diffusion depending on the laser vector potential and the average energy of the electron beam. The evolution of the electron energy distribution function can be also interpreted in terms of entropy. Neitz and Di Piazza have shown that the entropy of the electron beam increases when the quantum stochasticity dominates, and decreases when the energy spread is decreasing \cite{Piazza_qed_energyspread}.

 We will now compare the above findings with simulation results obtained using the OSIRIS-QED framework. The simulation setup is depicted in Fig. \ref{qedrr_cartoon} where an electron bunch is colliding head-on with a circularly polarised laser and emitting photons. The electrons are presented in their transverse momentum space as a function of the longitudinal spatial coordinate. The main characteristic of the quantum radiation emission is already visible in this figure: even though there is an average trend to emit and to lose more energy further into the laser pulse, the energy of an individual electron is subject to fluctuations due to stochastic nature of the quantum photon emission. 

To illustrate these features, we first present a set of simulations using two different electron bunches with mean energies of 0.5 GeV and 0.85 GeV. The bunches are initialised with a very small thermal momentum spread, equal in all transverse directions (the initial beam divergence is $p_\perp/p_\parallel\sim$ 0.2 mrad). The laser is modeled as a transverse plane wave with a temporal envelope function $f(t)$. The laser rise and fall sections have the same shape and duration $\tau_{rise}=\tau_{fall}=50.0\ \omega_0^{-1}$, while the duration of the flat part $\tau_{flat}$ is varied between 0.0 and 300.0 $\omega_0^{-1}$ with a step of 50.0 $\omega_0^{-1}$ (seven different total pulse durations $\tau=\tau_{flat}+(\tau_{rise}+\tau_{fall})/2$). The slope of the envelope function for $t<\tau_{rise}$ is defined as $f(t)=10(t/\tau_{rise})^3-15(t/\tau_{rise})^4+6(t/\tau_{rise})^5$, where $\tau_{rise}=50.0\ \omega_0^{-1}=26.6\  \mathrm{fs}$ and $\omega_0=1.88\times10^{15}$. The variable length middle section of the pulse had the laser vector potential always at the maximum value ($a_0=27$).  We shall stress that the interaction between the particles of the beam is negligible and since the laser field amplitude is uniform in the transverse directions, all the particles are subject to the same conditions. This corresponds to the approximation of beam transverse size much smaller than the laser spot size. Therefore, the large amount of PIC particles (about 1 million) provides us with a good statistical sample to study the evolution of the energy spectra of the electrons. The simulations are performed in 2D, where the box size was $500\times20\ c^2/\omega_0^2$, resolved with $5000\times200$ cells and the timestep $dt=0.04\ \omega_0^{-1}$ using 16 particles per cell. The simulation timestep is chosen such that $dt\ll W_{rad}^{-1}$, where $W_{rad}$ is the photon emission rate. 

The results of the first set of simulations are summarized in Fig. \ref{qedrr_spect} which shows the electron energy spectra after the interaction with the laser. All the spectra, as expected, are wider than the  quasi-monoenergetic initial distribution. The striking fact is that after interaction with longer lasers, the final energy spread of the electron beam is narrower than after interacting with shorter lasers. This hints that in case of longer interaction, a "turning point" predicted by the theory with properties given by Eq. (\ref{eqsigmacp}) must exist. After the initial increase of the width of the spectrum due to the quantum nature of the radiation process, the "turning point" indicates the time at which the width reduces anew as predicted by classical radiation reaction \cite{First_PRL}.

\begin{figure}
\centering
\includegraphics[width=1.0\textwidth]{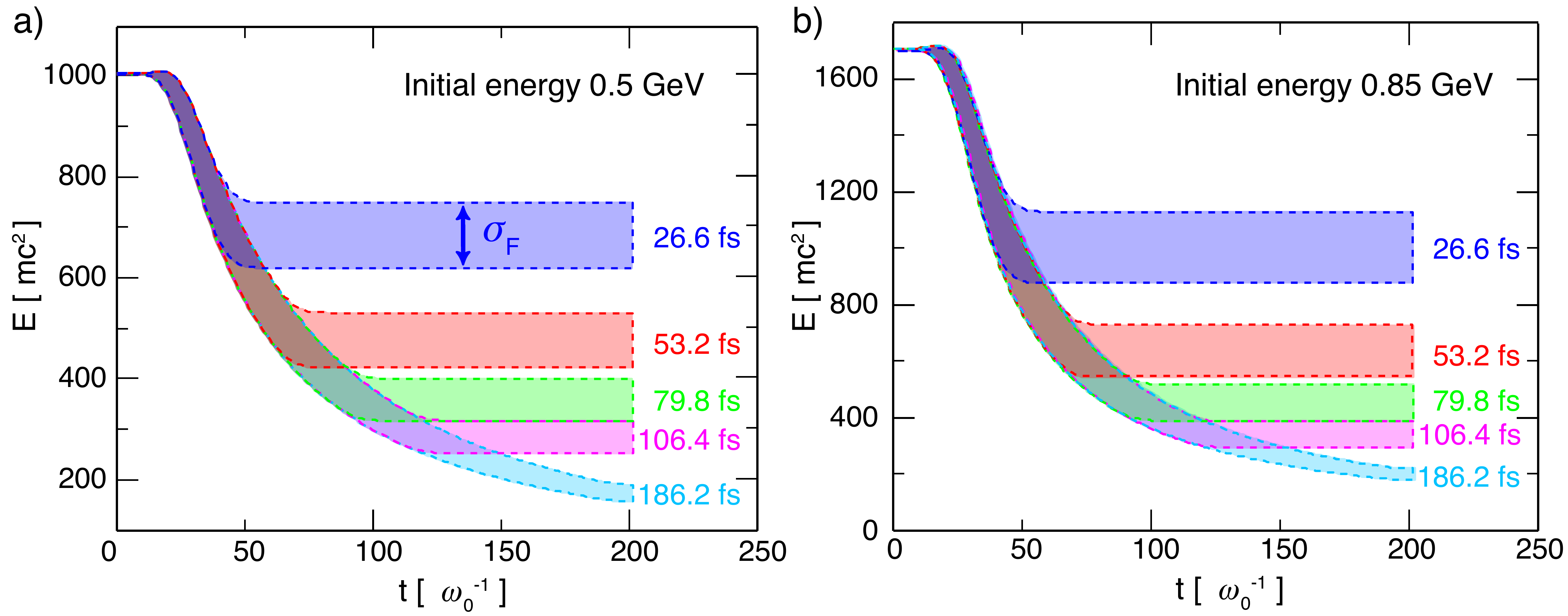}
\caption{Electron average energy evolution vs. time with standard deviation as a measure of the energy spread. The electron initial energy is a) 0.5 GeV and b) 0.85 GeV. Different colours denote different laser durations.} 
\label{qedrrg1700}
\end{figure}

To investigate this further, the evolution of the beam energy with the spread $\overline{\gamma(t)}\pm \sigma(t)/2$ as a function of time for several examples is shown in Fig. \ref{qedrrg1700}.  
Without quantum stochasticity or radiation reaction, both the electron energy and $\sigma(t)$ would remain the same throughout the whole interaction. With only classical radiation reaction, both the average energy and $\sigma(t)$ would reduce with time. The analysis of the beam spectral width evolution during the time of the interaction confirms that the first quantum effect is to broaden the spectrum due to quantum stochasticity. If the laser is short enough, the spectrum stays broad (this is in agreement with Ref. \cite{Piazza_qed_energyspread}). However, if the laser is longer, then there is a specific point in time where the spread starts decreasing due to the classical drift of the electron energy distribution function. 

\begin{figure}
\centering
\includegraphics[width=1.0\textwidth]{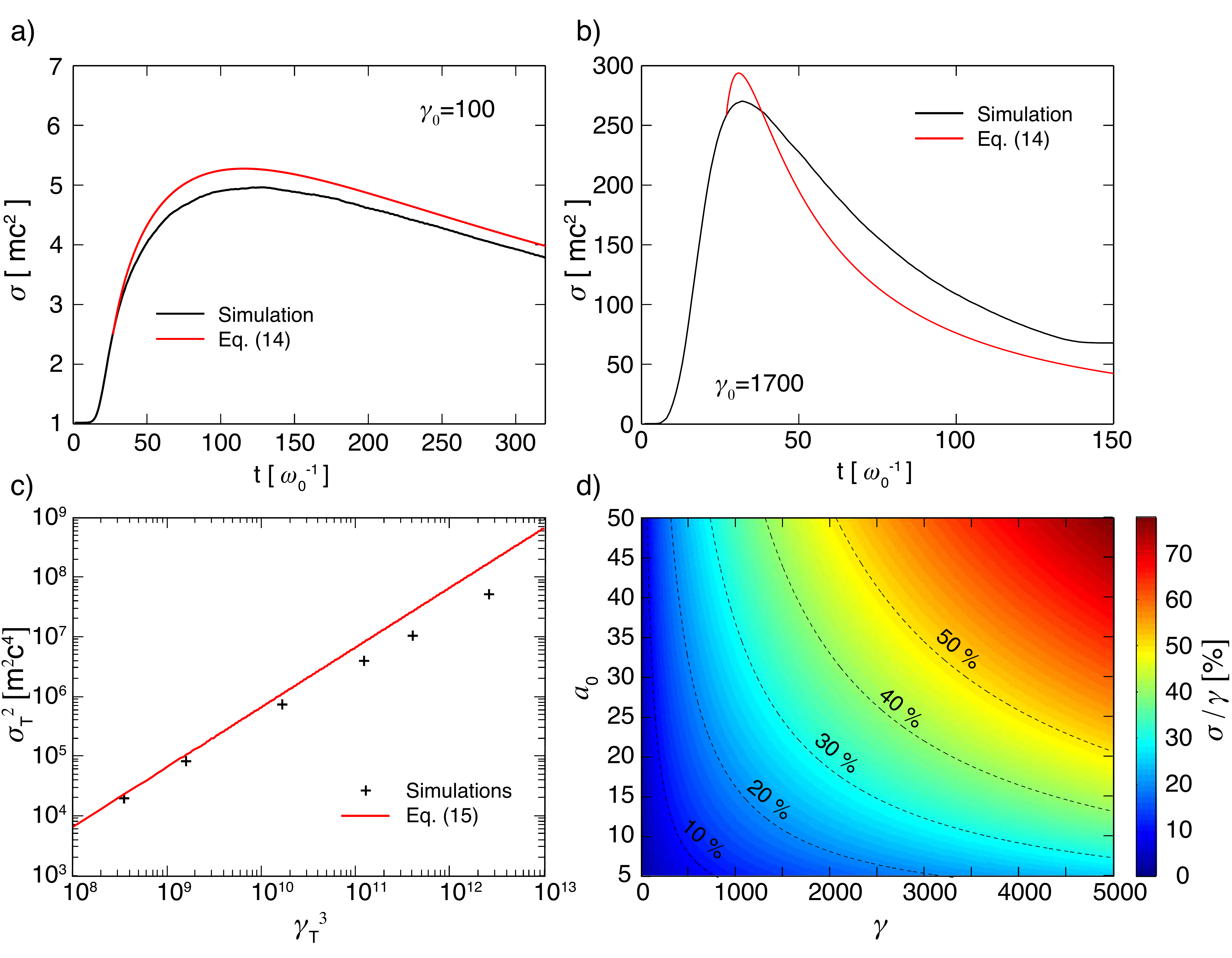}
\caption{ Electron beam energy spread.  a) Evolution vs. time for an electron beam starting at $\gamma_0=100$. The black line represents the data from the simulation, and the red line comes from numerical integration of Eq. (\ref{realdsig})  b) Same as in a) but for an electron beam with initial energy $\gamma_0=1700$.  c) Standard deviation of the distribution at the "turning point" as a function of $\gamma_T^3$. The line is determined by Eq. (\ref{eqsigmacp}) for $a_0=27$, while the points are the simulation data corresponding to the time at which the energy spread reaches its maximum ($d\sigma/dt=0$). d) Maximum attainable energy spread through diffusion depending on the energy of the particle and the normalised vector potential of the wave (in percentages).  } 
\label{max_sigma}
\end{figure}

A second set of simulations is performed by varying the electron beam initial energy, using a laser pulse similar to the ones described previously ($a_0=27$, $\tau_{\mathrm{rise}}=\tau_{\mathrm{fall}}=50 ~\omega_0^{-1}$, $\tau_{\mathrm{flat}}=600 ~\omega_0^{-1}$), with the same simulation box and resolution.  We would like to compare the predictions of Eq. (\ref{realdsig}) with the simulation results in a regime with $\chi_e\ll 1$, in a wave with a constant amplitude that allows for direct integration of Eq. (\ref{realdsig}). The black line in Fig. \ref{max_sigma} a) shows the evolution of the energy spread from a simulation with $\gamma=100$, where $\chi_e\simeq 0.01$. We are interested only in the constant amplitude section, so we select $t=27~\omega_0^{-1}$ as the new ``initial time''($\tau_{\mathrm{rise}}/2 + \tau_{\mathrm{beam}}$, where $\tau_{beam}=2\ \omega_0^{-1}$ is the electron beam "duration"). Therefore, $\sigma_0=\sigma(t=27 ~\omega_0^{-1} )$ and $\gamma_0=\gamma(t=27 ~\omega_0^{-1})$. The numerical integration of Eq. (\ref{realdsig}) is computed from this new ``initial time'' in order to compare with the simulation results. The panel a) of Fig. \ref{max_sigma} shows that Eq. (\ref{realdsig}) gives a result in good agreement with the simulation. Even in the case of a higher $\gamma_0=1700$, which corresponds to $\chi_e\simeq 0.2$, the integration of Eq. (\ref{realdsig}) provides a reasonable agreement (Fig. \ref{max_sigma} b)).

In Fig. \ref{max_sigma} c) we show $\sigma_T^2$ as a function of $\gamma_T^3$ at the "turning point" for several simulations starting at different average electron energies. All the simulations are performed with $a_0=27$  and the "turning point" is located within the constant amplitude section of the pulse. For particles with lower energies (and therefore with lower $\chi_e$), the ``turning point'' is well identified by Eq. (\ref{eqsigmacp}). For higher energies, the $\chi_e$ parameter is close to one and the electron energy spread is high, which makes the simulation results depart from the prediction of Eq. (\ref{eqsigmacp}). However, the value obtained in the simulations is always lower than the predicted value.  This confirms that the upper limit on the electron energy spread increase through diffusion as a function of $a_0$ and $\gamma$ can be estimated using Eqs. (\ref{eqsigmacp}) and (\ref{eqsigmalp}). The predictions of Eq. (\ref{eqsigmacp}) are shown in Fig. \ref{max_sigma} d) for a range of different values of the laser intensities and electron energies. 

\begin{figure}
\centering
\includegraphics[width=1.0\textwidth]{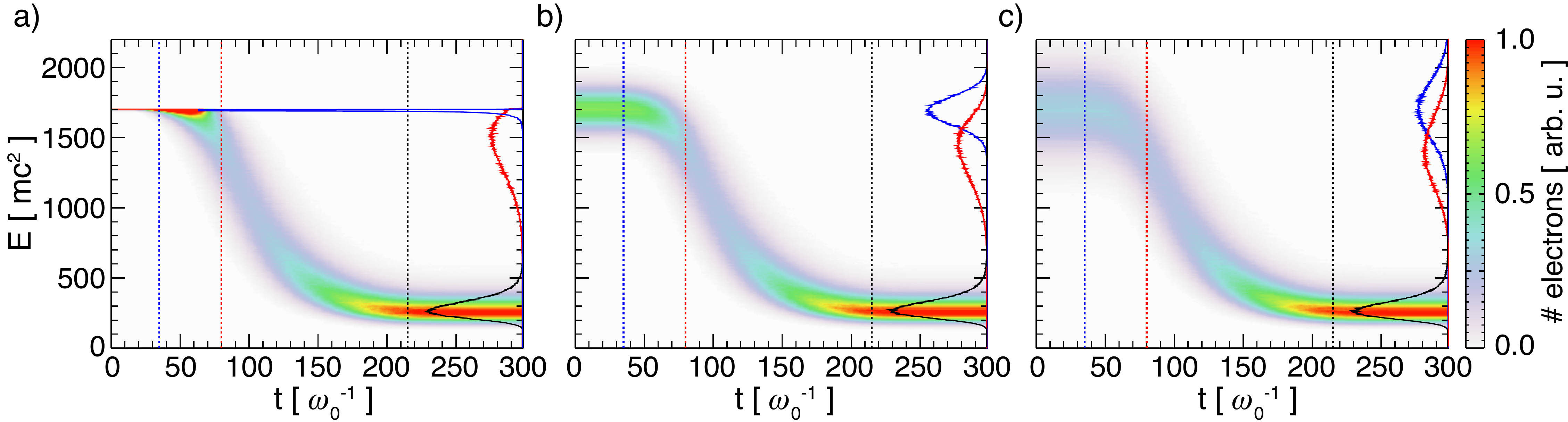}
\caption{ Electron beam energy spectrum evolution in time in a 150 fs laser pulse. Average initial energy of the electron beam is $\gamma_0=1700$, with the initial energy spread of  a) $\sigma_0=2$, b) $\sigma_0=100$ and c) $\sigma_0=200$. The lineouts represent the electron spectra at times $t=35~\omega_0^{-1}$ (blue), $t=80~\omega_0^{-1}$ (red) and $t=215~\omega_0^{-1}$ (black). After the shutdown of the laser, $\sigma=67.2$ for a)-c).} 
\label{ELI_example}
\end{figure}

Let us comment on the underlying physics involved in this behaviour. The differential probability rate of photon emission given by Eq. (\ref{compton_rate}) depends on the $\chi_e$ parameter in such manner that electrons with higher $\chi_e$ emit on average a larger fraction of their energy than the electrons with low $\chi_e$. This is what leads to the classical-like shrinking of the electron beam energy distribution in addition to the average energy drift towards a lower value. Moreover, the photons in the nonlinear Compton regime are emitted according to a distribution, such that electrons in identical conditions can radiate photons of different energy; this leads to a diffusion in the electron distribution function. These two tendencies compete, and the drift effect becomes dominant if the energy spread is wide enough ($\sigma\gtrsim \sigma_T$). On the contrary, if the initial electron energy spread is very low, the diffusion process dominates. This is illustrated in Fig. \ref{ELI_example} that shows the temporal evolution of electron energy spectra in a Gaussian laser pulse with duration $\tau=150~ \mathrm{fs}$ and peak vector potential $a_0=27$ (a pulse like this will be available, for instance, in ELI Beamlines \cite{ELI}). An electron beam is initialized with $\gamma_0=1700$, and we varied the value of $\sigma_0$. The first value $\sigma_0=2$ (Fig.  \ref{ELI_example} a) corresponds to an electron beam energy spread of 0.1 \% that is still a challenge for experiments. The second $\sigma_0=100$ in Fig. \ref{ELI_example} b) corresponds to a 6 \% energy spread, that has already been achieved with GeV-class electron beams in state-of-the-art laser facilities \cite{Leemans_4gev}. Third value $\sigma_0=200$ in Fig. \ref{ELI_example} c) corresponds to 12 \% energy spread, and this is routinely achievable with laser wakefield acceleration in the present day laboratory conditions.

\begin{figure}
\centering
\includegraphics[width=1.0\textwidth]{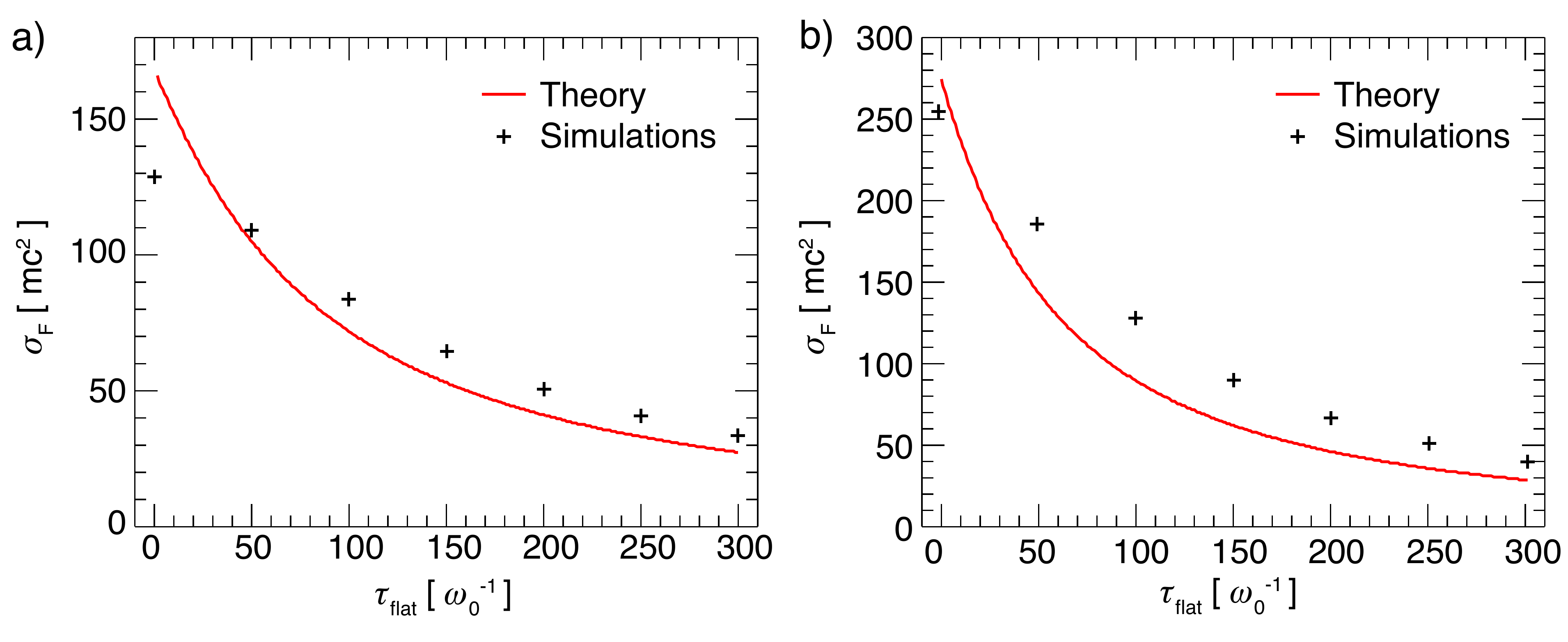}
\caption{a), b) Final electron energy spread for different laser durations and initial electron energy of a) 0.5 GeV and b) 0.85 GeV. Red dashed line represents the numerical integration of Eq. (\ref{realdsig}), blue line is given by Eq. (\ref{explicit_final_sigma}), while points are obtained directly from the simulations. c), d) Predictions for the final electron energy spread after interacting with a laser pulse at $I=2\times 10^{21}~\mathrm{W/cm^2}$ as a function of initial electron energy. Two typical intense laser durations of 30 fs and 100 fs are considered.} 
\label{fig_final_sigma}
\end{figure}

It is clear from Fig. \ref{ELI_example}  that starting from a very narrow distribution, the diffusion appears to be faster than for an initial wider one. All the three examples finally converge to the same electron energy distribution function. We can take advantage of this to calculate the expected electron energy spread at the end of the interaction. We perform the numerical integration of Eq. (\ref{realdsig}) assuming that the initial energy spread $\sigma_0$ is equal to the maximum attainable energy spread through diffusion $\sigma_T$ determined by Eq. (\ref{eqsigmacp}) for a given $a_0$ and $\overline{\gamma_0}$. The result obtained in this way is valid for all $\sigma_0<\sigma_T$, as long as the total time of interaction is much longer than the typical emission time $T_{total}\gg W_{rad}^{-1}$. The expected final energy spread obtained through numerical integration of Eq. (\ref{realdsig}) is compared with the simulation results in Fig. \ref{fig_final_sigma} a), b) for all the different laser pulse durations considered in the first set of simulations. As expected, the agreement is better for longer laser pulses and lower average electron energies, but the order of the expected energy spread is well predicted in all cases (maximum error is about 30\%). 

For guiding future experiments, it would be beneficial to have an explicit expression where one could insert the initial electron and laser parameters, and estimate the final energy spread of the electron beam that could be measured directly on a spectrometer. It is possible to attain such an approximate expression for laser pulses with a symmetrical, Gaussian-like temporal profile. We assume that the balance point $d\sigma=0$ has been reached before the centre of the pulse. This assumption is reasonable for beams with $\sigma_0\sim\sigma_T$, and for beams with $\sigma_0 \ll \sigma_T$ provided that there is enough time for the diffusion to act and increase $\sigma_0$ to the same order as $\sigma_T$. This is verified when the pulse duration satisfies $\tau_\mathrm{fwhm} >1/(8\gamma_0 \alpha_\mathrm{rr})$. If the pulse is shorter, the stochastic effects dominate over the drift, and the final energy spread $\sigma_F$ can be directly estimated using Eq. (\ref{width_qed_an}). Similarly, for $\sigma_0\gg \sigma_T$, the drift dominates and we can approximate $\sigma_F$ through Eq. (\ref{sigmaclass}). The most difficult case is, therefore, when $\sigma_0\sim \sigma_T$, as all the terms in Eq. (\ref{realdsig}) are of the same order, which renders the equation unintegrable. However, we can estimate an upper boundary for $\sigma_F$ by assuming that at the central point of the laser (at the point of peak intensity) the electron beam is close to the balance between the drift and diffusion, i.e.  $\sigma_M^2\approx (2.4/\lambda[\mu \mathrm{m}])\times10^{-6}\gamma_M^3a_0$, where $\gamma_M$ is the average Lorentz factor of the electron beam in the central laser point. As $\gamma_M$ is easy to calculate (see \cite{ ThomasRR, First_PRL, Bulanov_LLLAD, Tikhonchuk, emittance_decrease}), we can retrieve an explicit expression for $\sigma_M$ as a function of laser intensity and duration, and initial electron energy. Beyond this point, the energy spread slowly decreases, and the final electron energy spread $\sigma_F$ is smaller than $\sigma_M$. This yields
\begin{equation}\label{explicit_final_sigma}
\sigma_F^2\lesssim1.455\times10^{-4} \sqrt{I_{22}} \frac{\gamma_0^3}{\left( 1+6.12\times 10^{-5}\gamma_0~ I_{22}~ \tau_0[\mathrm{fs}] \right)^3},
\end{equation} 
where $I_{22}= I\left[10^{22}~\mathrm{W/cm^2}\right]$ and $a_0=0.855 \sqrt{I[10^{18}~\mathrm{W/cm^2}]} \lambda[\mu \mathrm{m}]$ for linear polarisation and $a_0=  0.855 \sqrt{I[10^{18}~\mathrm{W/cm^2}]} \lambda[\mu \mathrm{m}]/\sqrt{2}$ for circular polarisation. It is worth noting that the result presented in Eq. (\ref{explicit_final_sigma}) does not depend on the laser polarisation, but solely on intensity and duration. 

Figure \ref{fig_final_sigma} a), b) shows the estimate given by Eq. (\ref{explicit_final_sigma}) compared with the simulation results. Even though the lasers in our simulations are not Gaussian, we obtain a satisfactory agreement for the same $\tau_\mathrm{fwhm}$. Panels c) and d) show the predictions for the final energy spread according to Eq. (\ref{explicit_final_sigma}) for electron beams starting at different initial energies after interacting with a 30 fs and a 100 fs laser of $2\times10^{21}~ \mathrm{W/cm^2}$ intensity. These laser durations are to be available in the near-future laser facilities such as ELI \cite{ELI}, so there is a possibility to verify this model in the next few years.

\section{Electron beam divergence}

\begin{figure}
\centering
\includegraphics[width=1.0\textwidth]{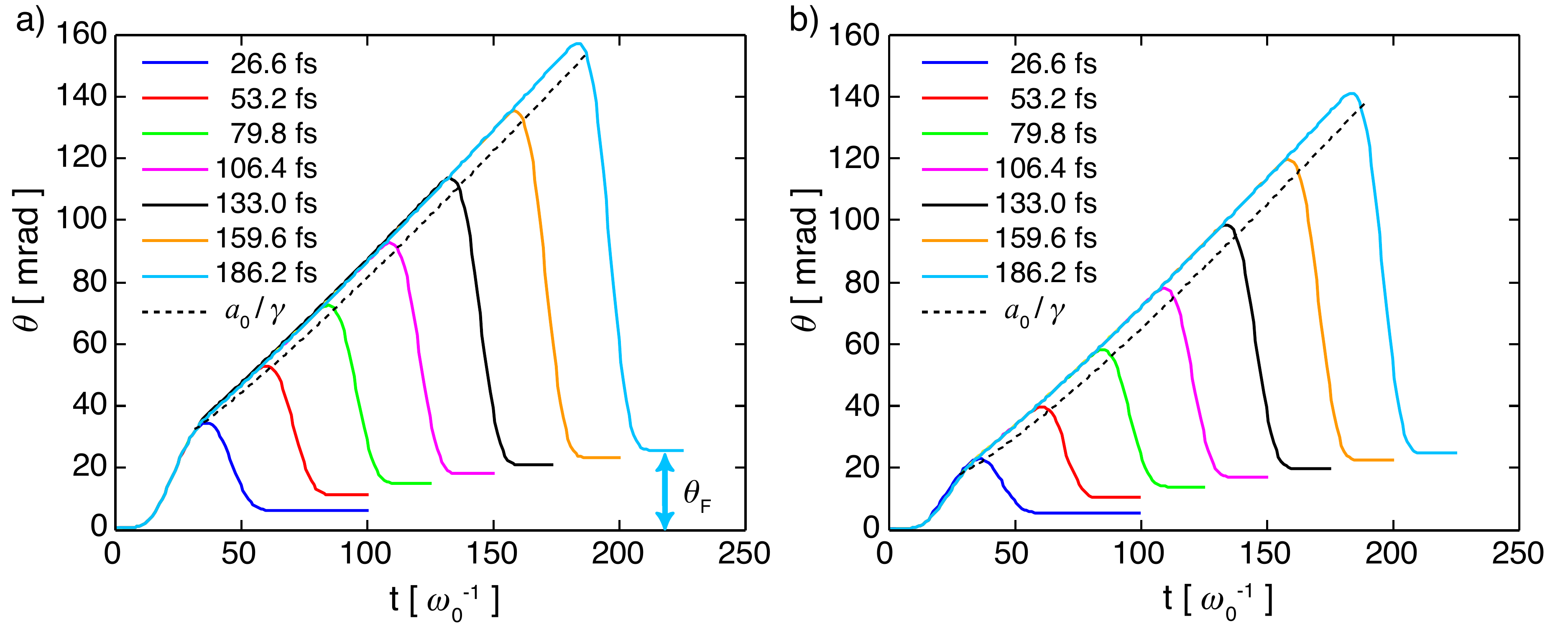}
\caption{Electron beam divergence vs. time. a) Electron beam initial energy is 0.5 GeV. b) Electron beam initial energy is 0.85 GeV.} 
\label{qedrr_angle}
\end{figure}

In addition to the electron energy spread, we can also evaluate the impact of the laser interaction on the electron beam divergence. We define the weighted average of the deflection angle from the main propagation direction as
\begin{equation}\label{avg_angle_def}
\overline{\tan \theta}=\frac{\sum_{i=1} ^N q_i \left( \frac{p_\perp}{p_{\parallel}} \right)_i}{\sum_{i=1} ^N q_i},
\end{equation} 
where $N$ is the total number of simulation particles, $q_i$ is the charge weight of the $i$-th particle, and $(p_\perp/p_\parallel)_i$ is the ratio of the transverse to the longitudinal momentum with respect to the direction of laser propagation. For small angles, $\tan \theta \simeq \theta$,  and the average divergence shown in Fig. \ref{qedrr_angle} is determined with this approximation (the error is less than 1 mrad). 

Figure \ref{qedrr_angle} shows the evolution of the electron bunch divergence as time progresses. Classically, the radiation reaction leads to momentum phasespace contraction proportionally in transverse and longitudinal direction. According to the analytical solution for trajectory of a relativistic electron in an intense plane wave \cite{Piazza_solutionLL}, on average, all momentum components and electron energy are reduced by a same factor due to radiation reaction. The angle between the particle momentum and the laser propagation direction is therefore approximately the same before and after the interaction, provided that the laser has a slowly varying temporal envelope compared to the laser period. However, during the interaction with the laser, the electron has an additional oscillatory component of the transverse momentum, whose amplitude depends only on the laser intensity ($p_\perp^*\simeq a_0mc$). It is worth noting that the oscillatory component of the $p_\perp$ given by the laser is the same with and without radiation reaction, as it depends only on the normalised vector potential of the wave $a_0$ (c. f. Supplementary material \cite{Supplemental_info}). 

In our case, the particles are counter-propagating with the laser, so the initial $p_{\perp0}=0$. The only transverse momentum of such a particle is therefore $p_\perp^*$. In the field of a circularly polarised wave, the transverse momentum vector rotates in the plane perpendicular to the laser propagation direction, while its magnitude remains constant $p_\perp=p_\perp^*\simeq a_0mc$. 

As the initial electron energy is on the order of a 0.5-1.0 GeV, we take $\gamma_0 \gg a_0$ and the parallel momentum becomes $p_\parallel=m c\sqrt{\gamma^2-a_0^2-1}\simeq  \gamma mc$. As a result, the average angle that a single electron makes with the direction of laser propagation is approximately  $\theta\simeq a_0/\gamma$. Without radiation reaction, this angle would stay the same during the constant amplitude section of the laser, as there would be no change in $\gamma$. However, Eq. (\ref{dgdg_help}) shows that with radiation reaction the electron Lorentz factor decreases through $\gamma=\gamma_0/(1+2\alpha_\mathrm{rr} \gamma_0 t)$ and the angle is then expected to increase linearly as a function of time:
\begin{equation}
\theta\simeq\frac{a_0}{\gamma_0}(1+2\alpha_\mathrm{rr} \gamma_0 t).
\end{equation}

The dashed lines in Fig. \ref{qedrr_angle} show the average expected angle $a_0/\gamma$ during the constant amplitude part of the pulse, where $\gamma$ is taken as the average relativistic factor of the electron bunch and $a_0=27$. We observe a similar trend with the simulation data, which indicates that the average divergence increase due to radiation emission in the constant amplitude region of the laser envelope is well explained by the semi-classical approach. However, there is a slight difference between the simulation data and the expected $a_0/\gamma$ which increases over time. 

After the interaction has finished, the electron beam has a residual divergence on the order of $\theta_\mathrm{F}\sim 10~ \mathrm{mrad}$ which is larger than the initial divergence on the 0.2 mrad level. Semi-classical radiation reaction predicts the final divergence to be approximately equal to its initial value. In QED, we expect the final  $\overline{p_\perp}=0$, but the divergence defined by Eq. (\ref{avg_angle_def}) can have a larger value than initial if there is a wider particle distribution function in transverse momentum space.  

\begin{figure}
\centering
\includegraphics[width=1.0\textwidth]{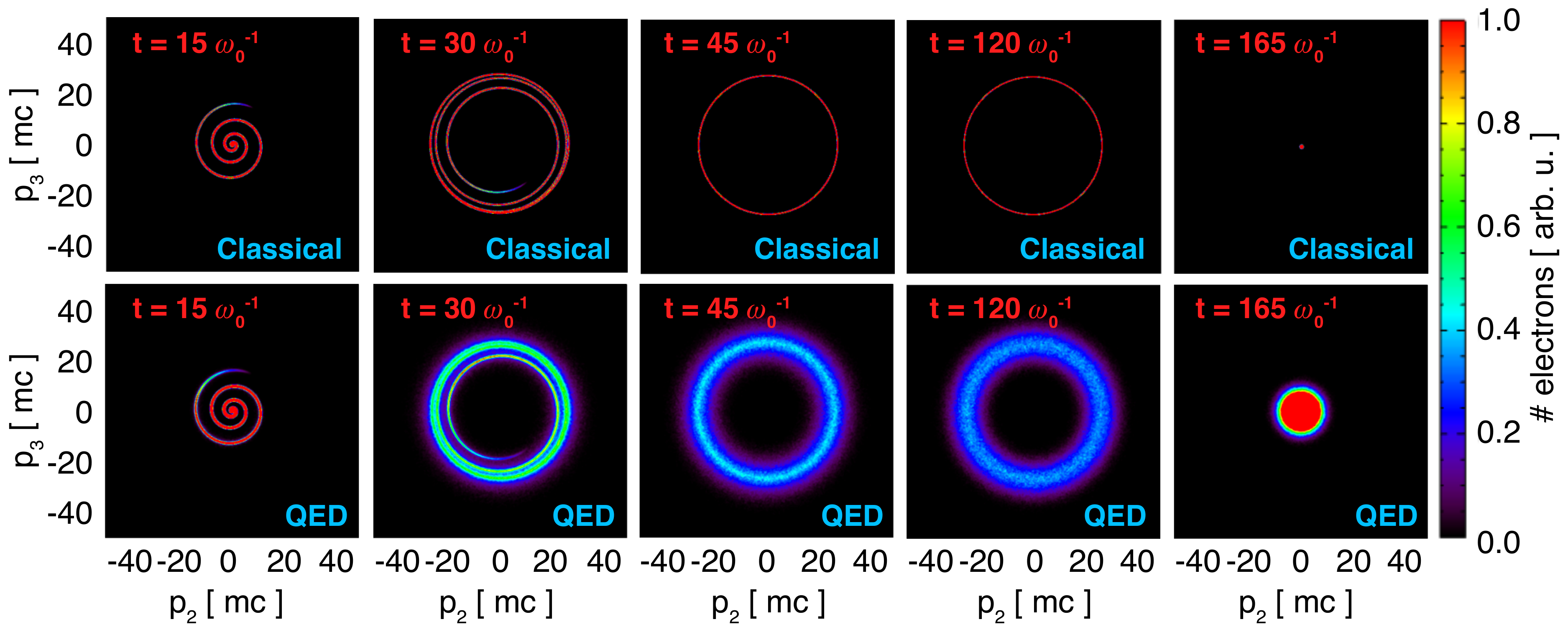}
\caption{Transverse momentum space $p_2-p_3$ at different times. Classical and QED radiation reaction give different final transverse momentum spread.} 
\label{qedrr_p2p3}
\end{figure}

To ascertain the origin of this effect, we have examined the transverse momentum space at different times (see Fig. \ref{qedrr_p2p3}). Initially the electron beam has a narrow momentum spread. During the plane wave stage the average transverse momentum is indeed around the predicted value $p_\perp\simeq a_0mc$. Howbeit, the QED simulations show the existence of a momentum spread around the average value that increases with time. This is consistent with the transverse momentum spreading previously reported in ref. \cite{Harvey_transverse_momentum}. The variation around the average angle as defined in Eq. (\ref{avg_angle_def}) during the interaction with the constant-amplitude section of the laser can be approximately related to the variation in energy:
\begin{equation}
\Delta \theta\simeq \frac{a_0}{\gamma^2}\Delta\gamma.
\end{equation}
This variation persists and finally becomes the net beam divergence when the laser shuts down: $\theta_\mathrm{F}\simeq \sqrt{2/\pi} \left(a_0/\gamma_\mathrm{F}^2\right)\sigma_F$. As $\gamma_F$ converges to a lower value for a longer interaction time, and $\sigma$ is from Eq. (\ref{eqsigmacp}) approximately proportional to $\gamma^{3/2}$, we can then conclude that the width of the final angular spread increases slowly with the length of the interaction (as seen in Fig. \ref{qedrr_angle}).

\section{Conclusions}
In classical radiation reaction, the energy loss of a single electron depends on its initial energy. For an electron beam, the main effects are the decrease in its mean energy and reduction of the energy distribution width. When QED effects are taken into account, the intrinsic stochastic nature of photon emission leads to diffusion in the energy distribution around the mean value which would translate into the increase of the energy spread of the beam. Therefore, in the general scenario, there is a competition between these two tendencies. If we allow a long enough interaction time, there is a point when the diffusion in momentum, intrinsically quantum, is balanced by the energy width reduction, associated with the classical regime. Beyond this point, the energy spread only decreases. This allows for estimating the maximal attainable energy spread through diffusion for a set of initial parameters $\gamma_0$ and $\sigma_0$. We have estimated this limit (and confirmed it with numerical simulations), which has further allowed us to predict analytically the final electron energy spread, which is the relevant quantity to be measured in experiments.  

The average divergence of the electron beam during the laser interaction is well-described by the classical radiation reaction. However, we have observed that the electron distribution function in momentum space has a certain spread around the average value that increases with the interaction time. This spread persists after the interaction is shut down and leads to a residual divergence of the electron beam that can be estimated analytically through its connection with the electron energy distribution function.

The control of beam properties is of relevance for all near future laser facilities that will operate at high intensities, regardless if they are aimed at optimising particle acceleration, radiation sources or fundamental research. As the quantum spreading might discriminate between the measurable effects and those whose signatures are too small to be observed due to the width of the final distribution function, our findings are vital for the design of upcoming experiments. They are also valuable for numerous applications with specific beam quality requirements.

\section*{Acknowledgements}
This work is supported by the European Research Council (ERC-2010-AdG Grant 267841) and FCT (Portugal) Grants SFRH/BD/62137/2009 and SFRH/IF/01780/2013. Simulations were performed at Supermuc (Germany) under a PRACE Grant and at the Accelerates cluster (Lisbon, Portugal).

\section*{References}
\bibliographystyle{unsrt}
\bibliography{FP_qedrr.bbl}

\begin{thebibliography}{10}

\bibitem{ELI}
The {ELI} project, http://www.extreme-light-infrastructure.eu/.

\bibitem{ELI_design}
Bruno Le~Garrec, Stephane Sebban, Daniele Margarone, Martin Precek, Stefan
  Weber, Ondrej Klimo, Georg Korn, and Bedrich Rus.
\newblock Eli-beamlines: extreme light infrastructure science and technology
  with ultra-intense lasers, 2014.

\bibitem{HiPER}
{HiPER} project, http://www.hiper-laser.org/.

\bibitem{Gremillet}
M.~Lobet, E.~dHumieres, M.~Grech, C.~Ruyer, X.~Davoine, and L.~Gremillet.
\newblock Modeling of radiative and quantum electrodynamics effects in pic
  simulations of ultra-relativistic laser-plasma interaction.
\newblock {\em J. Phys. Conf. Ser.}, 688:012058, 2016.

\bibitem{Ridgers_quantumRRnew}
T.~G. Blackburn, C.~P. Ridgers, J.~G. Kirk, and A.~R. Bell.
\newblock Quantum radiation reaction in laserÐelectron-beam collisions.
\newblock {\em Phys. Rev. Lett.}, 112:015001, Jan 2014.

\bibitem{Sarri_ep}
G.~Sarri, K.~Poder, J.~M. Cole, W.~Schumaker, A.~Di~Piazza, B.~Reville,
  T.~Dzelzanis, D.~Doria, L.~A. Gizzi, G.~Grittani, S.~Kar, C.~H. Keitel,
  K.~Krushelnick, S.~Kuschel, S.~P.~D. Mangles, Z.~Najmudin, N.~Shukla, L.~O.
  Silva, D.~Symes, A.~G.~R. Thomas, M.~Vargas, J.~Vieira, and M.~Zepf.
\newblock Generation of neutral and high-density electron-positron pair plasmas
  in the laboratory.
\newblock {\em Nat. Commun.}, 46:6747, April 2015.

\bibitem{Bulanov_multiple_lasers}
S.~S. Bulanov, V.~D. Mur, N.~B. Narozhny, J.~Nees, and V.~S. Popov.
\newblock Multiple colliding electromagnetic pulses: A way to lower the
  threshold of ${e}^{+}{e}^{-}$ pair production from vacuum.
\newblock {\em Phys. Rev. Lett.}, 104:220404, Jun 2010.

\bibitem{Yoffe_evolution}
Samuel~R Yoffe, Yevgen Kravets, Adam Noble, and Dino~A Jaroszynski.
\newblock Longitudinal and transverse cooling of relativistic electron beams in
  intense laser pulses.
\newblock {\em New Journal of Physics}, 17(5):053025, 2015.

\bibitem{Elkina_rot}
N.~V. Elkina, A.~M. Fedotov, I.~Yu. Kostyukov, M.~V. Legkov, N.~B. Narozhny,
  E.~N. Nerush, and H.~Ruhl.
\newblock Qed cascades induced by circularly polarized laser fields.
\newblock {\em Phys. Rev. ST Accel. Beams}, 14:054401, May 2011.

\bibitem{Ridgers_solid}
C.~P. Ridgers, C.~S. Brady, R.~Duclous, J.~G. Kirk, K.~Bennett, T.~D. Arber,
  A.~P.~L. Robinson, and A.~R. Bell.
\newblock Dense electron-positron plasmas and ultraintense $\gamma${} rays from
  laser-irradiated solids.
\newblock {\em Phys. Rev. Lett.}, 108:165006, Apr 2012.

\bibitem{Nerush_laserlimit}
E.~N. Nerush, I.~Yu. Kostyukov, A.~M. Fedotov, N.~B. Narozhny, N.~V. Elkina,
  and H.~Ruhl.
\newblock Laser field absorption in self-generated electron-positron pair
  plasma.
\newblock {\em Phys. Rev. Lett.}, 106:035001, Jan 2011.

\bibitem{Bell_Kirk_MC}
R.~Duclous, J.~G. Kirk, and A.~R. Bell.
\newblock Monte carlo calculations of pair production in high-intensity
  laser-plasma interactions.
\newblock {\em Plasma Phys. Contr. F.}, 53(1):015009, 2011.

\bibitem{QED1}
C.~Bula, K.~T. McDonald, E.~J. Prebys, C.~Bamber, S.~Boege, T.~Kotseroglou,
  A.~C. Melissinos, D.~D. Meyerhofer, W.~Ragg, D.~L. Burke, R.~C. Field,
  G.~Horton-Smith, A.~C. Odian, J.~E. Spencer, D.~Walz, S.~C. Berridge, W.~M.
  Bugg, K.~Shmakov, and A.~W. Weidemann.
\newblock Observation of nonlinear effects in compton scattering.
\newblock {\em Phys. Rev. Lett.}, 76:3116--3119, Apr 1996.

\bibitem{QED2}
D.~L. Burke, R.~C. Field, G.~Horton-Smith, J.~E. Spencer, D.~Walz, S.~C.
  Berridge, W.~M. Bugg, K.~Shmakov, A.~W. Weidemann, C.~Bula, K.~T. McDonald,
  E.~J. Prebys, C.~Bamber, S.~J. Boege, T.~Koffas, T.~Kotseroglou, A.~C.
  Melissinos, D.~D. Meyerhofer, D.~A. Reis, and W.~Ragg.
\newblock Positron production in multiphoton light-by-light scattering.
\newblock {\em Phys. Rev. Lett.}, 79:1626--1629, Sep 1997.

\bibitem{QED3}
C.~Bamber, S.~J. Boege, T.~Koffas, T.~Kotseroglou, A.~C. Melissinos, D.~D.
  Meyerhofer, D.~A. Reis, W.~Ragg, C.~Bula, K.~T. McDonald, E.~J. Prebys, D.~L.
  Burke, R.~C. Field, G.~Horton-Smith, J.~E. Spencer, D.~Walz, S.~C. Berridge,
  W.~M. Bugg, K.~Shmakov, and A.~W. Weidemann.
\newblock Studies of nonlinear qed in collisions of 46.6 gev electrons with
  intense laser pulses.
\newblock {\em Phys. Rev. D}, 60:092004, Oct 1999.

\bibitem{Schwinger}
J.~Schwinger.
\newblock On gauge invariance and vacuum polarization.
\newblock {\em Phys. Rev.}, 82:664--679, Jun 1951.

\bibitem{Ritus_thesis}
V.~I. Ritus.
\newblock Quantum effects of the interaction of elementary particles with an
  intense electromagnetic field.
\newblock {\em Journal of Soviet Laser Research}, 6(5):497--617, 1985.

\bibitem{kogaPOP}
J.~Koga, T.~Z. Esirkepov, and S.~V. Bulanov.
\newblock Nonlinear thomson scattering in the strong radiation damping regime.
\newblock {\em Phys. Plasmas}, 12(9):093106, 2005.

\bibitem{capturePiazza}
A.~Di~Piazza, K.~Z. Hatsagortsyan, and C.~H. Keitel.
\newblock Strong signatures of radiation reaction below the radiation-dominated
  regime.
\newblock {\em Phys. Rev. Lett.}, 102:254802, Jun 2009.

\bibitem{ThomasRR}
A.~G.~R. Thomas, C.~P. Ridgers, S.~S. Bulanov, B.~J. Griffin, and S.~P.~D.
  Mangles.
\newblock Strong radiation-damping effects in a gamma-ray source generated by
  the interaction of a high-intensity laser with a wakefield-accelerated
  electron beam.
\newblock {\em Phys. Rev. X}, 2:041004, Oct 2012.

\bibitem{Keita_neweq}
K.~Seto, H.~Nagatomo, J.~Koga, and K.~Mima.
\newblock Equation of motion with radiation reaction in ultrarelativistic
  laser-electron interactions.
\newblock {\em Phys. Plasmas}, 18(12):123101, 2011.

\bibitem{RRinCascade_classical}
A.~Zhidkov, S.~Masuda, S.~S. Bulanov, J.~Koga, T.~Hosokai, and R.~Kodama.
\newblock Radiation reaction effects in cascade scattering of intense, tightly
  focused laser pulses by relativistic electrons: Classical approach.
\newblock {\em Phys. Rev. ST Accel. Beams}, 17:054001, May 2014.

\bibitem{Ruhl_threshold}
Y.~Hadad, L.~Labun, J.~Rafelski, N.~Elkina, C.~Klier, and H.~Ruhl.
\newblock Effects of radiation reaction in relativistic laser acceleration.
\newblock {\em Phys. Rev. D}, 82:096012, Nov 2010.

\bibitem{Leemans_4gev}
W.~P. Leemans, A.~J. Gonsalves, H.-S. Mao, K.~Nakamura, C.~Benedetti, C.~B.
  Schroeder, Cs. T\'oth, J.~Daniels, D.~E. Mittelberger, S.~S. Bulanov, J.-L.
  Vay, C.~G.~R. Geddes, and E.~Esarey.
\newblock Multi-gev electron beams from capillary-discharge-guided subpetawatt
  laser pulses in the self-trapping regime.
\newblock {\em Phys. Rev. Lett.}, 113:245002, Dec 2014.

\bibitem{First_PRL}
M.~Vranic, J.~L. Martins, J.~Vieira, R.~A. Fonseca, and L.~O. Silva.
\newblock All-optical radiation reaction at
  ${10}^{21}\mathrm{W}/{\mathrm{cm}}^{2}$.
\newblock {\em Phys. Rev. Lett.}, 113:134801, Sep 2014.

\bibitem{LLbook}
L.~D. {Landau} and E.~M. {Lifshitz}.
\newblock {\em {The Classical Theory of Fields}}.
\newblock Butterworth Heinemann, 1975.

\bibitem{LLfromQEDsoviet}
V.S. Krivitski and V.~N. Txytovich.
\newblock Average radiation-reaction force in quantum electrodynamics.
\newblock {\em Sov. Phys. Usp.}, 34:250--258, 1991.

\bibitem{spohn}
H.~Spohn.
\newblock The critical manifold of the lorentz-dirac equation.
\newblock {\em Europhys. Lett.}, 50(3):287, May 2000.

\bibitem{Spohnbook}
H.~{Spohn}.
\newblock {\em {Dynamics of charged particles and their radiation field}}.
\newblock Cambridge University Press, 2004.

\bibitem{Piazza_solutionLL}
A.~Di~Piazza.
\newblock Exact solution of the landau-lifshitz equation in a plane wave.
\newblock {\em Lett. Math. Phys.}, 83(3):305--313, 2008.

\bibitem{qed_class_rr}
A.~Ilderton and G.~Torgrimsson.
\newblock Radiation reaction in strong field qed.
\newblock {\em Physics Letters B}, 725(4Ð5):481 -- 486, 2013.

\bibitem{Vranic_ClassicalRR}
M.~Vranic, J.~L. Martins, R.~A. Fonseca, and L.~O. Silva.
\newblock Classical radiation reaction in particle-in-cell simulations.
\newblock {\em Comput. Phys. Commun.}, 204:141--151, 2016.

\bibitem{esarey_thompson}
E.~Esarey.
\newblock Laser cooling of electron beams via thomson scattering.
\newblock {\em Nucl. Instr. Meth. Phys. Res.}, 455(1):7 -- 14, 2000.

\bibitem{RR_in_fusion}
R.~D. Hazeltine and S.~M. Mahajan.
\newblock Radiation reaction in fusion plasmas.
\newblock {\em Phys. Rev. E}, 70:046407, Oct 2004.

\bibitem{Tamburini_NIMR}
M.~Tamburini, F.~Pegoraro, A.~Di Piazza, C.H. Keitel, T.V. Liseykina, and
  A.~Macchi.
\newblock Radiation reaction effects on electron nonlinear dynamics and ion
  acceleration in laserÐsolid interaction.
\newblock {\em Nuclear Instruments and Methods in Physics Research Section A:
  Accelerators, Spectrometers, Detectors and Associated Equipment}, 653(1):181
  -- 185, 2011.
\newblock Superstrong 2010.

\bibitem{Piazza_qed_energyspread}
N.~Neitz and A.~Di~Piazza.
\newblock Stochasticity effects in quantum radiation reaction.
\newblock {\em Phys. Rev. Lett.}, 111:054802, Aug 2013.

\bibitem{Piazza_QRR_FD}
A.~Di~Piazza, K.~Z. Hatsagortsyan, and C.~H. Keitel.
\newblock Quantum radiation reaction effects in multiphoton compton scattering.
\newblock {\em Phys. Rev. Lett.}, 105:220403, Nov 2010.

\bibitem{Volkov}
D.~M. Volkov.
\newblock On a class of solutions of the dirac equation.
\newblock {\em Z. Phys}, 94:250--260, 1935.

\bibitem{pair_rate1}
A.~I. Nikishov and V.~I. Ritus.
\newblock Pair production by a photon and photon emission by an electron in the
  field of ultra intense electromagnetic wave and in a constant field.
\newblock {\em Sov. Phys. JETP}, 25(6), 1967.

\bibitem{pair_rate2}
V.N. Baier and V.M. Katkov.
\newblock Quantum effects in magnetic bremsstrahlung.
\newblock {\em Phys. Lett. A}, 25(7):492 -- 493, 1967.

\bibitem{pair_rate3}
N.~P. Klepikov.
\newblock Emission of photons or electron-positron pairs in magnetic fields.
\newblock {\em Zhur. Esptl. i Teoret. Fiz.}, 26, 1954.

\bibitem{Erber}
T.~Erber.
\newblock High-energy electromagnetic conversion processes in intense magnetic
  fields.
\newblock {\em Rev. Mod. Phys.}, 38:626--659, Oct 1966.

\bibitem{NikishovRitus}
A.~I. Nikishov and V.~I. Ritus.
\newblock Quantum processes in the field of a plane electromagnetic wave and in
  a constant field.
\newblock {\em Sov. Phys. JETP}, 19:529--541, 1964.

\bibitem{Thomas_POP_2016}
T.~Grismayer, M.~Vranic, J.~L. Martins, R.~A. Fonseca, and L.~O. Silva.
\newblock Laser absorption via quantum electrodynamics cascades in counter
  propagating laser pulses.
\newblock {\em Phys. Plasmas}, 23:056706, 2016.

\bibitem{thomasQED}
T.~Grismayer, M.~Vranic, J.~L. Martins, R.~A. Fonseca, and L.~O. Silva.
\newblock Seeded qed cascades in counter propagating lasers.
\newblock {\em ArXiv}, 1511.07503, 2015.

\bibitem{OSIRIS}
R.~A. Fonseca, L.~O. Silva, F.~S. Tsung, V.~K. Decyk, W.~Lu, C.~Ren, W.~B.
  Mori, S.~Deng, S.~Lee, T.~Katsouleas, and J.~C. Adam.
\newblock {\em {OSIRIS: A three-dimensional, fully relativistic particle in
  cell code for modeling plasma based accelerators}}, volume 2331.
\newblock Springer Berlin / Heidelberg, 2002.

\bibitem{Gonoskov_schemesQEDPIC}
A.~Gonoskov, S.~Bastrakov, E.~Efimenko, A.~Ilderton, M.~Marklund, I.~Meyerov,
  A.~Muraviev, A.~Sergeev, I.~Surmin, and E.~Wallin.
\newblock Extended particle-in-cell schemes for physics in ultrastrong laser
  fields: Review and developments.
\newblock {\em Phys. Rev. E}, 92:023305, Aug 2015.

\bibitem{Vranic_merging}
M.~Vranic, T.~Grismayer, J.L. Martins, R.A. Fonseca, and L.O. Silva.
\newblock Particle merging algorithm for pic codes.
\newblock {\em Comput. Phys. Commun.}, 191:65--73, 2015.

\bibitem{LLphyskin}
E.~M. Lifshitz and L.~P. Pitaevskii.
\newblock {\em {Physical Kinetics}}.
\newblock Butterworth Heinemann, 1981.

\bibitem{FP-Planck}
M.~Planck.
\newblock An essay on statistical dynamics and its amplification in the quantum
  theory.
\newblock {\em Sitz.ber. Preu§. Akad}, 1:324--341, 1917.

\bibitem{FP-Fokker}
D.~A. Fokker.
\newblock Die mittlere energie rotierender elektrischer dipole im
  strahlungsfeld.
\newblock {\em Ann. Phys.}, 43:810--820, 1914.

\bibitem{Bulanov_LLLAD}
S.~V. Bulanov, T.~Z. Esirkepov, M.~Kando, J.~K. Koga, and S.~S. Bulanov.
\newblock Lorentz-abraham-dirac versus landau-lifshitz radiation friction force
  in the ultrarelativistic electron interaction with electromagnetic wave
  (exact solutions).
\newblock {\em Phys. Rev. E}, 84:056605, Nov 2011.

\bibitem{Tikhonchuk}
T.~Schlegel and V.~T. Tikhonchuk.
\newblock Classical radiation effects on relativistic electrons in ultraintense
  laser fields with circular polarization.
\newblock {\em New J. Phys}, 14(7):073034, 2012.

\bibitem{emittance_decrease}
E.~Esarey.
\newblock Laser cooling of electron beams via thomson scattering.
\newblock {\em Nucl. Instr. Meth. Phys. Res.}, 455(1):7 -- 14, 2000.

\bibitem{Supplemental_info}
Supplementary material.

\bibitem{Harvey_transverse_momentum}
D.~G. Green and C.~N. Harvey.
\newblock Transverse spreading of electrons in high-intensity laser fields.
\newblock {\em Phys. Rev. Lett.}, 112:164801, Apr 2014.

\end{thebibliography}



\end{document}